\documentclass[11pt,a4paper]{article}
\usepackage{amsmath,amssymb,amsfonts,bm}
\usepackage{graphicx}
\usepackage[top=1cm,body={16cm,23cm}]{geometry}

\begin{document}

\title{FROM UNINTEGRATED GLUON DISTRIBUTIONS \\
TO PARTICLE PRODUCTION\\
IN HADRONIC COLLISIONS AT HIGH ENERGIES
}

\author{A. SZCZUREK \\
Institute of Nuclear Physics \\
PL-31-342 Cracow, Poland,\\
University of Rzesz\'ow,\\
PL-35-959 Rzesz\'ow, Poland,\\
E-mail: antoni.szczurek@ifj.edu.pl }

\maketitle

\begin{abstract}
\noindent
The inclusive distributions of gluons and pions
for high-energy NN collisions are calculated.
The results for several unintegrated gluon distributions (UGD's)
from the literature are compared.  
We find huge differences in both rapidity and $p_t$
of gluons and $\pi$'s in NN collisions
for different models of UGD's.
The Karzeev-Levin UGD gives good description
of momentum distribution of charged hadrons at midrapidities.
We find that the gluonic mechanism discussed does not describe
the inclusive spectra of charged particles in the fragmentation
region.
\end{abstract}

%---------------------
\section{Introduction}
%---------------------

The recent results from RHIC (see e.g. \cite{RHIC}) have attracted
renewed interest in better understanding the dynamics of
particle production, not only in nuclear collisions.
Quite different approaches have been used
to describe the particle spectra from the nuclear collisions \cite{PHOBOS}.
The model in Ref.\cite{KL01} with an educated guess
for UGD describes surprisingly well
the whole charged particle rapidity distribution by means
of gluonic mechanisms only. Such a gluonic mechanism would lead to
the identical production of positively and negatively charged hadrons.
The recent results of the BRAHMS experiment \cite{BRAHMS} put into
question the successful description of Ref.\cite{KL01}.
In the light of this experiment, it becomes obvious that
the large rapidity regions have more complicated flavour structure.

I discuss the relation between UGD's in hadrons and
the inclusive momentum distribution
of particles produced in hadronic collisions.
The results obtained with different UGD's
\cite{KL01,AKMS94,GBW_glue,KMR,Blue95} are shown and compared.

%-----------------------------------
\section{Inclusive gluon production}
%-----------------------------------

At sufficiently high energy the cross section for inclusive
gluon production in $h_1 + h_2 \rightarrow g$ can be written
in terms of the UGD's ``in'' both colliding
hadrons \cite{GLR81}
\begin{equation}
\frac{d \sigma}{dy d^2 p_t} = \frac{16  N_c}{N_c^2 - 1}
\frac{1}{p_t^2}
\int
 \alpha_s(\Omega^2)
 {\cal F}_1(x_1,\kappa_1^2) {\cal F}_2(x_2,\kappa_2^2)
\delta(\vec{\kappa}_1+\vec{\kappa}_2 - \vec{p}_t)
\; d^2 \kappa_1 d^2 \kappa_2    \; .
\label{inclusive_glue0}
\end{equation}
Above ${\cal F}_1$ and ${\cal F}_2$ are UGD's
in hadron $h_1$ and $h_2$, respectively.
The longitudinal momentum fractions are fixed by kinematics:
$x_{1/2} = \frac{p_t}{\sqrt{s}} \cdot \exp(\pm y)$.
The argument of the running coupling constant is taken as
$\Omega^2 = \max(\kappa_1^2,\kappa_2^2,p_t^2)$.

\begin{figure}[thb]
\begin{center}
    \includegraphics[width=8cm]{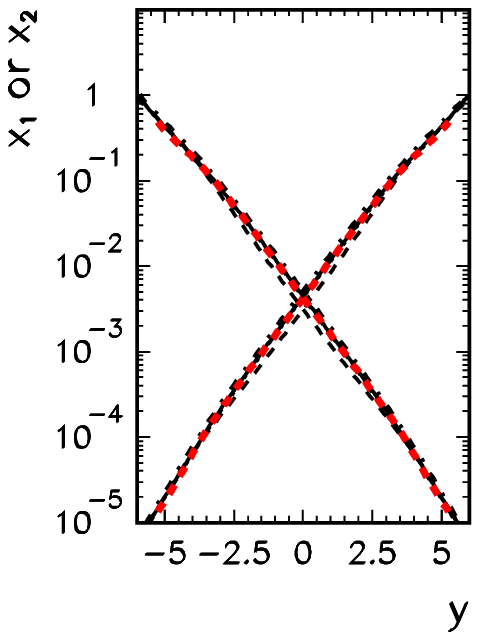}
\caption{
 $<x_1>$ and $<x_2>$
for $p_t >$ 0.5 GeV and at W = 200 GeV. 
\label{fig:x1_x2}
}
\end{center}
\end{figure}

Here I shall not discuss the distributions of ``produced'' gluons,
which can be found in \cite{Sz03}.
Instead I shall discuss what are typical values of $x_1$ and $x_2$
in the jet (particle) production.
Average value $<x_1>$ and $<x_2>$, shown in Fig.\ref{fig:x1_x2},
only weakly depend on the model of UGD.
For $y \sim$ 0 at the RHIC energy W = 200 GeV one tests
UGD's at $x_g$ = 10$^{-3}$ - 10$^{-2}$.
When $|y|$ grows one tests more and more asymmetric (in $x_1$ and $x_2$)
configurations. For large $|y|$ either $x_1$ is extremely
small ($x_1 <$ 10$^{-4}$) and $x_2 \rightarrow$ 1
or $x_1 \rightarrow$ 1 and $x_2$ is extremely small ($x_2 <$ 10$^{-4}$).
These are regions of gluon momentum fraction where the UGD's
is rather poorly known. The approximation used in obtaining
UGD's are valid certainly only for $x <$ 0.1.
In order to extrapolate the gluon distribution to
$x_g \rightarrow$ 1 I multiply
the gluon distributions from the previous section by a factor
$(1-x_g)^n$, where n = 5-7.

%---------------------------------------------------------------------

%---------------------------------------------------------------------

%---------------------------------------------
\section{From gluon to particle distributions}
%---------------------------------------------

In Ref.\cite{KL01} it was assumed, based on the concept
of local parton-hadron duality, that the rapidity distribution
of particles is identical to the rapidity distribution of gluons.
In the present approach I follow a different approach
which makes use of phenomenological fragmentation functions (FF's).
For our present exploratory study it seems
sufficient to assume $\theta_h = \theta_g$.
This is equivalent to $\eta_h = \eta_g = y_g$, where $\eta_h$ and
$\eta_g$ are hadron and gluon pseudorapitity, respectively. Then
\begin{equation}
y_g = \mathrm{arsinh} \left( \frac{m_{t,h}}{p_{t,h}} \sinh y_h \right)
\; ,
\label{yg_yh}
\end{equation}
where the transverse mass $m_{t,h} = \sqrt{m_h^2 + p_{t,h}^2}$.
In order to introduce phenomenological FF's
one has to define a new kinematical variable.
In accord with $e^+e^-$ and $e p$ collisions I define a 
quantity $z$ by the equation $E_h = z E_g$.
This leads to the relation
\begin{equation}
p_{t,g} = \frac{p_{t,h}}{z} J(m_{t,h},y_h) \; ,
\label{ptg_pth}
\end{equation}
where $J(m_{t,h},y_h)$ is given in Ref.\cite{Sz03}.
Now we can write the single particle distribution
in terms of the gluon distribution as follows
\begin{eqnarray}
\frac{d \sigma (\eta_h, p_{t,h})}{d \eta_h d^2 p_{t,h}} =
\int d y_g d^2 p_{t,g} \int 
dz \; D_{g \rightarrow h}(z,\mu_D^2) \\ \nonumber
\delta(y_g - \eta_h) \; 
\delta^2\left(\vec{p}_{t,h} - \frac{z \vec{p}_{t,g}}{J}\right)
\cdot \frac{d \sigma (y_g, p_{t,g})}{d y_g d^2 p_{t,g}} \; .
\label{from_gluons_to_particles}
\end{eqnarray}

In the present calculation I shall use only LO
FF's from \cite{BKK95}.

%--------------------------------------------------
\begin{figure}[thb]
\begin{center}
    \includegraphics[width=8cm]{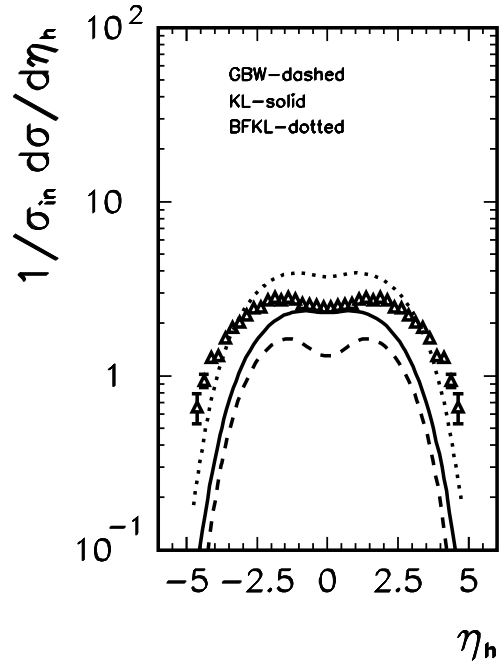}
\caption{
Charged-pion pseudrapidity distribution at W = 200 GeV
for different models of UGD's.
In this calculation $p_{t,h} >$ 0.2 GeV.
The experimental data of the UA5 collaboration are taken
from \cite{UA5_exp}.
\label{fig:eta_glue}
}
\end{center}
\end{figure}

%-----------------------------------------------------------------------

Let us analyze now how the results for pseudorapidity distributions
depend on the choice of the UGD.
In Fig.\ref{fig:eta_glue} I compare pseudorapidity distribution
of charged pions for different models of UGD's.
In this calculation FF from \cite{BKK95} has been used.

In contrast to Ref.\cite{KL01}, where the whole pseudorapidity
distribution, including fragmentation regions, has been well
described in an approach similar to the one presented here,
in the present approach pions produced from the fragmentation
of gluons in the $gg \rightarrow g$ mechanism populate only
midrapidity region,
leaving room for other mechanisms in the fragmentation regions.
These mechanisms involve quark/antiquark degrees of freedom
or leading protons among others. This strongly suggests
that the agreement of the result of the $gg \rightarrow g$
approach with the PHOBOS distributions \cite{PHOBOS} in
Ref.\cite{KL01} in the true fragmentation region is rather due to
approximations made in \cite{KL01} than due to correctness
of the reaction mechanism. In principle, this can be verified
experimentally at RHIC by measuring the $\pi^+ / \pi^-$ ratio
in proton-proton scattering as a function of (pseudo)rapidity
in possibly broad range.
The BRAHMS experiment can do it even with the existing apparatus. 
 
\begin{figure}[thb]
\begin{center}
    \includegraphics[width=8cm]{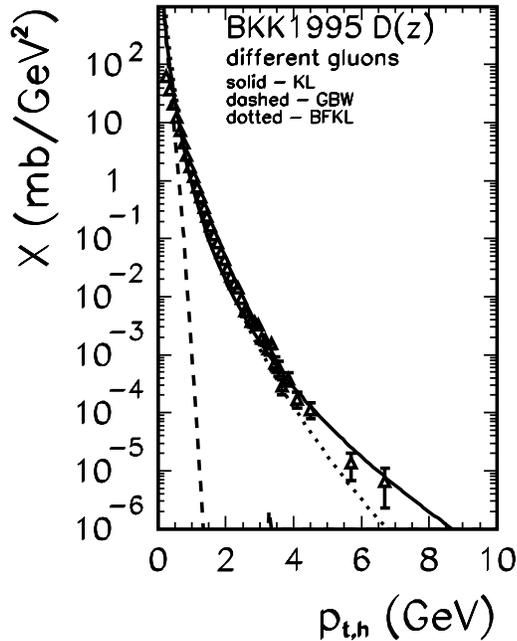}
\caption{
Transverse momentum distributions of charged pions at W = 200 GeV
for BKK1995 FF and different UGD's.
The experimental data are taken from \cite{UA1_exp}.
\label{fig:pt_glue}
}
\end{center}
\end{figure}

In Fig.\ref{fig:pt_glue} I compare the theoretical transverse
momentum distributions of charged pions obtained with
different gluon distributions with the UA1 collaboration data
\cite{UA1_exp}.
The best agreement is obtained with the
Karzeev-Levin gluon distribution. The distribution with
the GBW model is much too steep in comparison to experimental
data. This is probably due to neglecting QCD evolution
in \cite{GBW_glue}.

%--------------------------------------------------------------------------

%-----------------------------------------------------------------

%--------------------
\section{Conclusions}
%--------------------

I have calculated the inclusive distributions of gluons and
associated charged $\pi$'s in the NN collisions
through the $g g \rightarrow g$ mechanism
in the $k_t$-factorization approach. The results for several
UGD's proposed recently have been compared. The results, especially
$p_{t,h}$ distributions, obtained with different models of
UGD's differ considerably.

Contrary to a recent claim in Ref.\cite{KL01},
we have found that the gluonic mechanism discussed does not describe
the inclusive spectra of charged particles in the fragmentation
region, i.e. in the region of large (pseudo)rapidities for
any UGD from the literature.
Clearly the gluonic mechanism is not the only one.

Since the mechanism considered is not complete, it is not possible
at present to precisely verify different models of UGD's.
The existing UGD's lead
to the contributions which almost exhaust the strength
at midrapidities and leave room for other mechanisms
in the fragmentation regions. It seems that a measurement of
$p_t$ distributions of particles at RHIC should
be helpful to test better different UGD's.

%---------------------------------------------------------

\end{document}